\documentclass[12pt]{article}
\usepackage{graphicx,epsfig}

\def\fbi{\rm fb^{-1}}

\def\ccbar{c {\bar c}}
\def\lsim{\mathrel{\raise.3ex\hbox{$<$\kern-.75em\lower1ex\hbox{$\sim$}}}}
\def\gsim{\mathrel{\raise.3ex\hbox{$>$\kern-.75em\lower1ex\hbox{$\sim$}}}}
\newcommand{ \slashchar }[1]{\setbox0=\hbox{$#1$}   
   \dimen0=\wd0                                     
   \setbox1=\hbox{/} \dimen1=\wd1                   
   \ifdim\dimen0>\dimen1                            
      \rlap{\hbox to \dimen0{\hfil/\hfil}}          
      #1                                            
   \else                                            
      \rlap{\hbox to \dimen1{\hfil$#1$\hfil}}       
      /                                             
   \fi}                                             %

\def\ie{{\it i.e.}}

\def\gev{\,{\rm GeV}}

\def\to{\rightarrow}
\def\re{\rm Re}
\def\im{\rm Im}
\def\slash{\not\!}
\def\be{\begin{equation}}
\def\ee{\end{equation}}
\def\bea{\begin{eqnarray}}
\def\eea{\end{eqnarray}}
\def\bec{\begin{center}}
\def\eec{\end{center}}
\def\atversim#1#2{\lower0.7ex\vbox{\baselineskip\zatskip\lineskip\zatskip
  \lineskiplimit 0pt\ialign{$\matth#1\hfil##\hfil$\crcr#2\crcr\sim\crcr}}}

\renewcommand{\thefootnote}{\fnsymbol{footnote}}

\newcounter{appendixc}
\newcounter{subappendixc}[appendixc]
\newcounter{subsubappendixc}[subappendixc]

\renewcommand{\appendix}[1] {\vspace*{0.6cm}
        \refstepcounter{appendixc}
        \setcounter{figure}{0}
        \setcounter{table}{0}
        \setcounter{equation}{0}
        \renewcommand{\thefigure}{\Alph{appendixc}.\arabic{figure}}
        \renewcommand{\thetable}{\Alph{appendixc}.\arabic{table}}
        \renewcommand{\theappendixc}{\Alph{appendixc}}
        \renewcommand{\theequation}{\Alph{appendixc}.\arabic{equation}}
        \noindent{\bf Appendix \theappendixc #1}\par\vspace*{0.4cm}}

\begin{document}

\begin{titlepage}
\rightline{\vbox{\halign{&#\hfil\cr &KEK-TH-939 \cr &UCL-IPT-04-01 \cr
&hep-ph/0401246\cr
}}}
\vskip .5in
\begin{center}

{\Large\bf Inclusive $J/\psi$ Productions at $e^+ e^-$ Colliders}

\vskip .5in

\normalsize {\bf  K. Hagiwara}$^1$, {\bf  E. Kou}$^2$, 
{\bf  Z.-H. Lin}$^1$, 
{\bf C.-F. Qiao}$^{3,4}$, {\bf G.-H. Zhu}$^1$\\
\vskip .5cm

$^1$ Theory Group, KEK, Tsukuba, Ibaraki 305-0801, Japan\\
\vskip .3cm

$^2$ Institut de Physique Th\'{e}orique, Universit\'{e} Catholique de Louvain,\\
Chemin Cyclotron~2, B-1348 Louvain-la-Neuve, Belgium
\vskip .3cm

$^3$ CCAST(World Lab.), P.O. Box 8730, Beijing 100080, China\\
\vskip .3cm

$^4$ Dept. of Physics, Graduate School of the Chinese
Academy of Sciences,\\
YuQuan Road 19A, Beijing 100039, China
\vskip .5cm

\end{center}

\begin{abstract}\normalsize
Inclusive $J/\psi$ productions in $e^+ e^-$ annihilation is studied in the 
framework of NRQCD. We first review the leading-order calculations of the 
cross sections for $e^+ e^- \to J/\psi c \bar{c}$ and $e^+ e^- \to J/\psi g g$ 
and find that their ratio is about 1:1.5 at $\sqrt{s}\simeq 10$\gev .  
This result is in conflict with the current measurements by the Belle 
Collaboration, which finds that the process $e^+e^- \to J/\psi c\bar{c}$ 
accounts for about 2/3 of all the prompt $J/\psi$'s. We show that the 
discrepancy in the total rate as well as in the $J/\psi$ momentum 
distributions can be resolved by considering a large renormalization $K$ factor
($K\simeq 4$) for the $J/\psi c \bar {c}$ cross section and by taking into 
account collinear suppression in the end-point energy region of 
$J/\psi g g$ production. Detailed studies of the model predictions in terms 
of the density matrix are performed and various momentum and angular 
distributions are presented as functions of the $K$ factors. These 
distributions can be used to determine the normalizations of each subprocess 
provided that the production and decay angular distributions do not alter 
much by higher order corrections. 

\end{abstract}
\vspace{1cm} PACS number(s): 12.38.Bx, 12.39.Jh, 13.60.Le, 14.40.Lb

\renewcommand{\thefootnote}{\arabic{footnote}}
\end{titlepage}


\section{Introduction}
\par
As one of the simplest processes to investigate both perturbative and 
nonperturbative properties of quantum chromodynamics (QCD), charmonium 
production at various collision processes has stimulated a lot of interesting 
theoretical and experimental works. One of such developments in the past few 
years, called nonrelativistic quantum chromodynamics (NRQCD)~\cite{nrqcd} which
generalizes and improves the conventional color-singlet model (CSM), has 
provided a successful explanation of the CDF measurements of prompt $J/\psi$ 
and $\psi^{\prime}$ production at the Tevatron~\cite{cdf}. Within the framework
of NRQCD, the puzzle of $\psi$ productions in excess of the CSM prediction can 
be solved by introducing significant contributions from color-octet terms, 
which correspond to a gluon forming a $c \bar{c}$ pair in a color-octet state 
at short distances and then evolving at long distances into a color-singlet 
state along with other light quarks. To confirm the validity
of the color-octet mechanism (COM), Braaten and Chen first suggested that the  
inclusive $\psi$ productions through $e^+ e^-$ annihilation may provide an 
opportunity to observe the color-octet contributions to the cross section 
and angular distributions ~\cite{braaten-chen}. Subsequently many detailed
calculations have been performed in the literature
~\cite{cho,chao1,lee,chang,chao2,chao3,others}.
\par
Recently, BaBar and Belle Collaborations have published their experimental data
for prompt $J/\psi$ productions~\cite{babar, belle1, belle2}. 
Both measurements for the inclusive processes are dramatically 
larger than the leading-order prediction of the CSM. 
However, there is no obvious evidence in support of the existence of the 
color-octet state reported by Belle measurement, 
especially in the upper end-point region of $J/\psi$ momentum 
distributions where the significant color-octet signal is predicted by previous analysis. 
Although higher order corrections may significantly soften the hard $J/\psi$ momentum 
spectrum of the COM~\cite{FLM03}, it is the following Belle observation 
\bea\label{rate}
\sigma(e^+ e^- \to J/\psi c {\bar c})/\sigma(e^+ e^- \to J/\psi X) =
0.59^{+0.15}_{-0.13}\pm 0.12=0.59\pm 0.18, 
\eea
which prompts us to reconsider the normalization of the $J/\psi c\bar{c}$ production 
in CSM. Even higher ratio, 
\bea\label{rate2}
\sigma(e^+ e^- \to J/\psi c {\bar c})/\sigma(e^+ e^- \to J/\psi X) =
0.67\pm 0.12, 
\eea
has been reported by Belle as a preliminary result 
based on $86.7 \fbi$ data set~\cite{belle3}. 
Since it is unlikely that COM gives the ratio larger than a quarter, we may conclude 
that the COM contribution is sub-dominant in the process $e^+e^- \to J/\psi X$ at 
$\sqrt{s}=10$\gev. 
\par
In this report, we assume that the COM contribution is negligible for 
prompt $J/\psi$ productions, and study if the CSM predictions for 
$e^+e^-\to J/\psi c\bar{c}$ and $e^+e^-\to J/\psi gg$ can be made consistent 
by introducing the renormalization $K$ factors for the total cross sections, 
and by considering the softening of 
the $J/\psi$ momentum spectrum in the $J/\psi gg$ process. We give predictions 
for the $J/\psi$ production and decay angular distributions, which will be useful to test 
our assumptions and to measure the $K$ factors and the momentum distributions of 
individual subprocesses in future precision experiments. 
\par
We note here some other evidences that may support our assumption indirectly.
The exclusive cross sections for the double-charmonium productions 
such as $e^+ e^- \to J/\psi \eta_c$, measured by Belle, is also one order of 
magnitude larger than the leading-order prediction~\cite{belle2}. 
Since in the exclusive processes, the color-octet contribution is negligible,  
a large $K$ factor for the $e^+e^-\to J/\psi \eta_c$ production amplitude in the 
CSM may be necessary to explain 
the experimental data, as has been illustrated in Ref.~\cite{hagiwara-kou-qiao}.   
\par
Lately some authors have calculated electromagnetic 
contributions and the relativistic correction~\cite{braaten-lee}, 
and also considered the possibility of mis-detecting 
the QED process $e^+ e^- \to J/\psi J/\psi$ at Belle~\cite{bbl,luchinsky}. 
However, since the considered subprocesses have the same order of 
magnitude of the  original tree-level cross section in CSM, they are  not sufficient
to provide an explanation of the large enhancement of the $J/\psi c\bar{c}$ cross section. 
In the recent report~\cite{belle4}, Belle collaboration showed that they are 
capable of distinguishing $J/\psi$ from $\eta_c$ and that they observe no evidence 
of $e^+e^-\to J/\psi J/\psi$ yet.  
\par
Another evidence of possible suppression of COM comes from radiative $\Upsilon$ decays. 
At the amplitude level,
$\Upsilon \to \gamma X$ is quite similar to inclusive $J/\psi$ productions 
except for the heavy quark masses. Recently,
radiative $\Upsilon$ decays observe by CLEO~\cite{cleo} has been studied 
in the framework of the combination of NRQCD and the soft-collinear effective theory (SCET)
~\cite{bauer,fleming}. 
The result shows that a good fit in the end-point region of the photon energy 
spectrum is obtained only when the color-octet matrix elements are set to zero. 
The same may or may not apply for prompt $J/\psi$ production in $e^+e^-$ collisions, 
but we can  
take into account the collinear suppression effect at the 
end-point region in $e^+ e^- \to J/\psi gg$ process, which is very similar to that of 
the $\Upsilon \to \gamma gg$ in the CSM.  
\par
In Section 2, we give the leading-order calculations of the processes
$e^+ e^- \to J/\psi c {\bar c}$ and $e^+ e^- \to J/\psi gg$ and correct some
mistakes in literatures. 
The production of $J/\psi c {\bar c}$ through two virtual photons
is also considered and the fractions for transversely and longitudinally polarized
$J/\psi$ are presented.
In Section 3, we extract the collinear suppression effect 
in $e^+e^- \to J/\psi gg$ production from the photon momentum spectrum of 
 $\Upsilon$ radiative decays, and find that the 
observed $J/\psi$ momentum spectrum and the $J/\psi c\bar{c}$ fraction 
Eq.(\ref{rate2}) can be reproduced by the CSM if we introduce a large $K$ 
factor of $K\simeq 4$ for 
the $J/\psi c\bar{c}$ cross section while the $J/\psi gg$ process does not need a 
$K$ factor significantly different from unity.  
In Section 4, we present various distributions for the inclusive $J/\psi$
productions in the different momentum regions and study the sensitivity 
of the $K$ factor in terms of several production and decay angular asymmetries. 
In Section 5, We discuss the connection between the inclusive process and 
the exclusive double-charmonium production. 
Finally, our conclusions are given in Section 6.
In Appendix A, we present technical details of the scheme which we developed
to calculate the helicity amplitudes of NRQCD processes by using the HELAS codes
~\cite{helas, madgraph}.
\section{The Leading-order Calculations} 
\par
In the color-singlet picture, three production modes are involved in the 
inclusive $J/\psi$ production processes, as shown in Fig.~1 and Fig.~2. 
We refer them to the QCD $J/\psi c {\bar c}$ production (Fig.~1(a)), 
the QED $J/\psi c {\bar c}$ production (Fig.~1(b),(c),(d)) and 
the $J/\psi gg$ production (Fig.~2).
Other QED processes such as $e^+ e^- \to J/\psi \gamma^*$ where the $\gamma^*$
forms a lepton or quark pair (except for the charm pair) have been regarded as
the background and removed experimentally~\cite{belle1}.
The $J/\psi gg$ production was initially investigated twenty years ago and was
taken as the most important production mode in comparison with the 
color-octet production in the old-fashioned "color-evaporation model"
~\cite{keung-kuhn}. The calculations on the QCD $J/\psi c {\bar c}$ production 
have been carried out in Refs.~\cite{cho,chao1,lee} and shows that the cross
section is comparatively smaller than that of $J/\psi gg$ production.
In the case of the $t$-channel QED $J/\psi c {\bar c}$ production (Fig. 1(b)), 
Ref.~\cite{chao3} shows that the cross section is only 1/6 of the QCD one 
and the modest effects are expected. Here in spite of a suppression factor 
$\alpha_{EM}^2/\alpha_s^2$ compared to the QCD $J/\psi c {\bar c}$ production,
the $t$-channel diagrams of the QED $J/\psi c {\bar c}$ process should
be taken into account due to the enhancement of powers of $\sqrt{s}/2m_c$,
which is studied in detail in Refs.~\cite{chang,bbl,chao3}.
According to the Furry's theorem, the density matrix of the interferences 
between $s$- and $t$- channel $J/\psi c {\bar c}$ processes (including the $s$-
channel QCD diagrams) vanishes when $c {\bar c}$ angular distributions in the 
$c {\bar c}$ rest frame are integrated out. 
We also calculate the corrections from $s$-channel QED productions
(the detailed procedures of our calculations will be shown later)
\bea\label{qedinter}
\frac{|a+c|^2-|a|^2}{|a|^2}&=&-3.0\%+3.5\times10^{-3}, \nonumber \\
\frac{|a+d|^2-|a|^2}{|a|^2}&=&2.0\%+1.0\times10^{-4},
\eea
where the symbols $|a|^2,|c|^2,|d|^2$ represent the contributions to the total 
cross section from diagrams Fig. 1(a), (c) and (d), respectively.
The first values on the right-hand side of Eq.~(\ref{qedinter}) represent
the corrections from the interference terms 
$2\re(ac^*)/|a|^2$ and $2\re(ad^*)/|a|^2$,
while the second values represent the direct corrections $|c|^2/|a|^2$ and 
$|d|^2/|a|^2$.
We note that the interference between the diagrams (a) and (c) is negative, 
which is consistent with the QED corrections to the exclusive process
$e^+ e^- \to J/\psi \eta_c$~\cite{braaten-lee}.
As expected, the sum of the above corrections from the $s$-channel 
QED processes contributes to the total $J/\psi c{\bar c}$ cross section
destructively by about 1\%, which is sufficiently small and can be neglected 
safely. Therefore, in the rest of the paper, we refer only to the
$t$-channel QED $J\psi c{\bar c}$ production of Fig1. (b) as the QED process.
\begin{figure}
\centerline{\epsfysize 5. truein \epsfbox{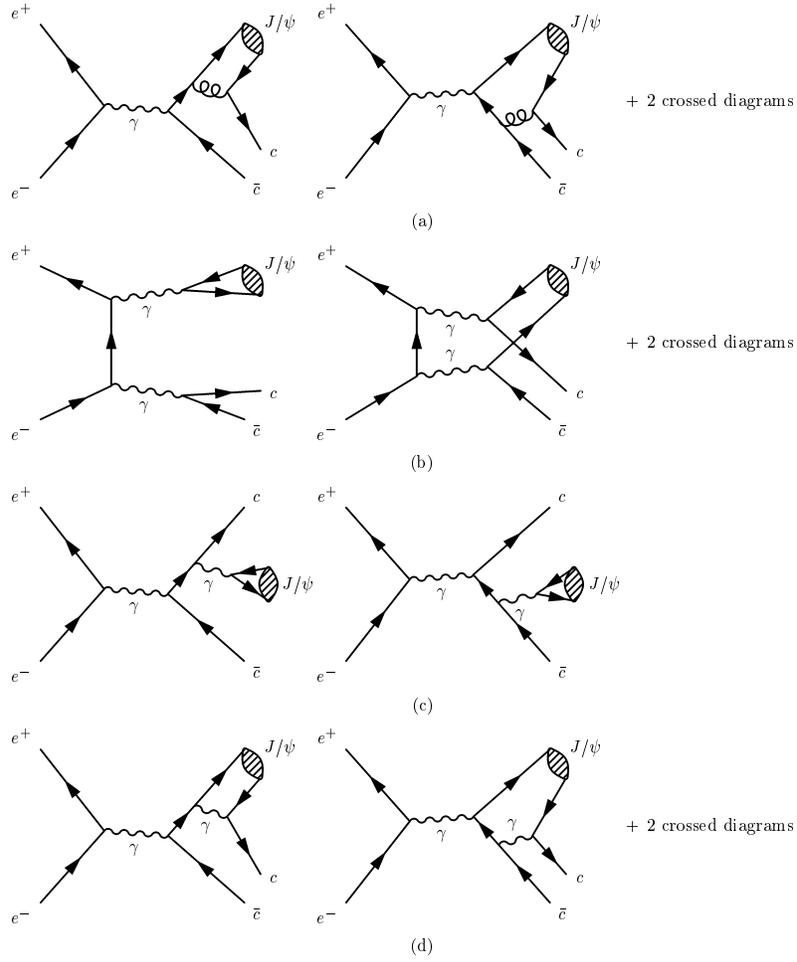}}
\caption{Feynman diagrams for the $J/\psi c {\bar c}$ productions 
from $e^+ + e^-$ annihilation: 
(a) the QCD production,
(b) the $t$-channel QED production,
(c) and (d) the $s$-channel QED productions. }
\end{figure}
\begin{figure}
\centerline{\epsfysize 1.25 truein \epsfbox{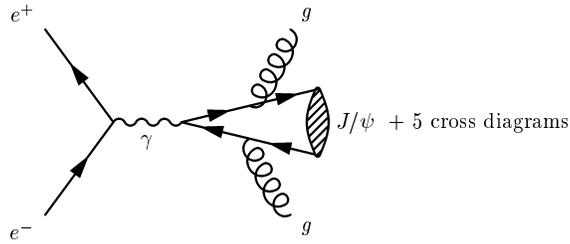}}
\caption{Feynman diagrams for 
the $J/\psi gg$ production from $e^+ + e^-$ annihilation. 
}
\end{figure}
\par
It has been noted that the color-octet process 
$e^+ e^- \to [c{\bar c}]_8 g \to J/\psi g$ exhibits 
a remarkably different property  
with all of the above three color-singlet productions. This process
gives rise to a hard spectrum where the $J/\psi$ momentum is almost maximal
because of its two-body final state~\cite{braaten-chen}. This has been a
crucial feature used in experiments to distinguish the color-octet 
contributions. As mentioned in the last section, no such 
signature has been detected by BaBar~\cite{babar} and 
Belle~\cite{belle1, belle2}. Recent investigation
reveals that the color-octet contribution to the $J/\psi$ spectrum can be 
broaden significantly  by the large perturbative corrections and enhanced 
nonperturbative effects~\cite{FLM03}. In this sense, one cannot rule out 
the color-octet contribution from current experimental data yet. 
However, as concluded in Ref.~\cite{FLM03}, due to the unknown 
color-octet shape functions for the $J/\psi$ production, 
the dominant $J/\psi c \bar{c}$ cross section must be understood accurately 
before extracting the color-octet contributions from the data.
\par
We now present our leading-order calculations. 
The traditional approach for $S$-wave productions 
is either calculating the helicity amplitudes
by using a covariant projection formalism in the CSM, 
or evaluating the squared amplitudes directly with the
help of the optical theorem in the general NRQCD factorization 
framework. To the lowest order in the power expansion in
the relativistic velocity $v$ of the heavy quark and anti-quark, 
the two methods are identical with each other.
In this paper, we develop a much simpler numerical method for 
the amplitude calculations. First we apply the program MadGraph
~\cite{madgraph} to generate the parton-level helicity amplitudes for 
the processes $e^+ e^- \to c {\bar c} c {\bar c}$ and 
$e^+ e^- \to c {\bar c} gg$, and then combine a color-singlet $c {\bar c}$ 
pair to be a $J/\psi$ meson by using the HELAS subroutines~\cite{helas}. 
We leave the detailed procedure for constructing the helicity amplitudes
for NRQCD processes in the appendix.
Finally the integration over the phase space of the squared
amplitudes are performed by the Monte Carlo program BASES~\cite{bases}.
This approach avoids lengthy trace computations and allows us to obtain various
production and decay angular distributions from the density matrices that
can easily be obtained from the helicity amplitudes.
\par
In our numerical analysis we use the input parameter values as follows:
the center-of-mass energy $\sqrt{s}=10.58$ GeV, $m_c=1.5$ GeV,
$\alpha_s(2m_c)=0.258$, and $\alpha_{EM}(2m_c)=1/129.6$.
The radial wave function for $J/\psi$ at the origin $R(0)$ is measured
through the leptonic decay width $\Gamma(J/\psi\to e^+e^-)$,
\bea\label{wave}
|R(0)|^2=\frac{9M^2_{J/\psi}}{16\alpha_{EM}^2}\Gamma(J/\psi\to e^+e^-)=0.447~{\rm GeV^3},
\eea
where $\Gamma(J/\psi\to e^+e^-)=5.26\times 10^{-6}\gev$, and $M_{J/\psi}\simeq 
2 m_c$.
Here we adopt the $R(0)$ value obtained from the above leading-order formula, 
rather than from the next-to-leading-order one~\cite{braaten-lee}
 or from some potential models
~\cite{quigg}, in order to define the renormalization $K$ factor of 
the $J/\psi$ production cross section unanimously. 
\par
With the above parameters, the leading-order cross sections for the inclusive 
$J/\psi$ processes are given as
\bea\label{lead}
\sigma_{cc}^{QCD}&=&0.0897~{\rm pb~~~
for~the~QCD~{\it J}/\psi c {\bar c}~production,}\nonumber \\
\sigma_{cc}^{QED}&=&0.0156~{\rm pb~~~
for~the~QED~{\it J}/\psi c {\bar c}~production,}\nonumber \\
\sigma_{gg}&=&0.162~{\rm pb~~~
for~the~{\it J}/\psi gg~production.} 
\eea
The ratio for the configurations of $J/\psi c {\bar c}$ and $J/\psi gg$,
\ie $(\sigma_{cc}^{QCD}+\sigma_{cc}^{QED})/\sigma_{gg}$, is about 1:1.5,
while $\sigma_{cc}^{QCD}/\sigma_{gg}=1:1.8$, which
are much larger than the previous results given in Refs.~\cite{cho,chao1,lee}.
For $\sigma_{cc}^{QCD}$, choosing the same input values used
in Refs.~\cite{cho,chao1,lee}, our result agrees with Ref.~\cite{lee} but
disagrees with that in Ref.~\cite{chao1}
and is three times larger than that in Ref.~\cite{cho}.
This missing factor 3 was also pointed out in Ref.~\cite{lee}.
In the case of $\sigma_{gg}$, our result is smaller than those
in Refs.~\cite{cho,lee} by about a factor of 2, but is consistent
with that in Refs.~\cite{keung-kuhn}. In addition, our result 
for $\sigma_{cc}^{QED}$ agrees with that in Ref.~\cite{chao3}.
Both $\sigma_{cc}^{QCD}$ and $\sigma_{gg}$ contribute to 
the total inclusive cross section at $O(\alpha_s^2 \alpha_{EM}^2)$,
while $\sigma_{cc}^{QED}$ contributes at $O(\alpha_{EM}^4)$.
The QED cross section is enhanced by powers of $\sqrt{s}/2m_c$.
As we shall see, the contribution from the QED $J/\psi c {\bar c}$ production
affects the angular distributions significantly 
at larger $J/\psi$ momenta.
\par
We also examine the consistency of our numerical results
with the fragmentation approximation
in the high energy limit; $\sqrt{s}\gg 2m_c$.
In Ref.~\cite{fragmentation}, the charm quark fragmentation function into 
$J/\psi$ has been defined as 
\bea
{\cal D}_{c\to J/\psi}(x)=\lim_{s\to\infty}
~\frac{1}{2~\sigma(e^+ e^- \to c {\bar c})}
~\frac{d\sigma^{QCD}}{dx}(e^+ e^-\to J/\psi c {\bar c})
\eea
and is found to be
\bea\label{frag}
{\cal D}_{c\to J/\psi}(x)=\frac{8}{27\pi}\alpha_s(2m_c)^2 
\frac{|R(0)|^2}{m_c^3}
\frac{x(1-x)^2(16-32x+72x^2-32x^3+5x^4)}{(2-x)^6}
\eea
with $x=2E_{J/\psi}/\sqrt{s}$ ($E_{J/\psi}$ is the $J/\psi$ energy).
We confirm the above result numerically and find that the deviation between 
the fragmentation formula (\ref{frag}) and the exact differential 
cross section decreases as $m_c/\sqrt{s}$ at high energies. 
The difference is about 6.2\% at $\sqrt{s}=50$ GeV and 2.4\% 
at $\sqrt{s}=100$ GeV.
\par
In Fig. 3, we show the momentum and angular distributions for the
three types of $J/\psi$ productions. Throughout the paper, we adopt
a dimensionless variable $z$ for the $J/\psi$ momentum distributions.
The relation between $z$ and the $J/\psi$ momentum $P_{J/\psi}$ is given
by $z=P_{J/\psi}/P_{J/\psi}^{max}$, where $P_{J/\psi}^{max}$ denotes
the maximum value of the $J/\psi$ momentum in the inclusive $J/\psi$ 
processes, namely, $P_{J/\psi}^{max}=(s-M_{J/\psi}^2)/2\sqrt{s}\simeq 4.86$ GeV.
For $e^+ e^- \to J/\psi c {\bar c}$, $P_{J/\psi}$ cannot achieve 
$P_{J/\psi}^{max}$ due to the kinematic constraint. The largest
$P_{J/\psi}$ is $\sqrt{s/4-4 m_c^2}$ which is about 4.36 GeV, corresponding 
to $z\simeq 0.90$ at $\sqrt{s}=10.58$ GeV. In the case of angular distributions,
we define $\theta$ as the opening angle between the produced $J/\psi$ momentum 
and the electron beam. Both $z$ and $\theta$ are defined in the $e^+ e^-$ 
collision center-of-mass frame.
\par
From Fig. 3(a) and (b), we observe that although the QED $J/\psi c {\bar c}$
cross section is much smaller than the QCD one, its effects 
in the high momentum region and at large $|\cos\theta|$ cannot be neglected. 
This is due to the characteristics of the photon fragmentation diagrams where 
one virtual photon fragments into a charmonium and another one evolves into 
a $c {\bar c}$ pair (see the left-most Feynman diagram in Fig. 1(b)).
The photon fragmentation configurations make up 71\% of the QED cross section,
while the rest comes from the interference terms and non-fragmentation terms.
In the end-point region $0.9<z<1$, only the $J/\psi gg$ mode is allowed.

The differential cross sections for the QCD $J/\psi c {\bar c}$ and the 
$J/\psi gg$ productions are restricted by unitarity, parity, and angular 
momentum considerations and can be parameterized to a simple form~\cite{cho}
\bea\label{cos}
\frac{d^2 \sigma}{dz d\cos\theta}=S(z)[1+\alpha(z)\cos^2\theta],
\eea
where the angular coefficient $\alpha(z)$ is generally limited in 
the interval $-1 \leq \alpha(z) \leq 1$. The angular distributions of the QCD 
$J/\psi c {\bar c}$ process (denoted by solid curves) in Fig. 3(b)
gives a positive $\alpha$ ($\alpha \sim 0.46$), corresponding to the dominance
of transverse $J/\psi$ mesons, while the flat shape of the $J/\psi gg$
angular distribution (dotted lines) gives $\alpha \sim 0$, corresponding to
a large fraction of the longitudinally polarized $J/\psi$ 
($\sigma_L / \sigma_T \sim 1.8$).  
These features can be confirmed in Fig. 3(c) and (d), where we show
the fraction of transversally polarized $J/\psi$ as function of $z$ and
$\cos\theta$, respectively.
In contrast to the above cosine-square behavior in Eq. (\ref{cos}), 
the differential cross section for the QED $J/\psi c {\bar c}$ production 
has strong enhancement near $|\cos\theta|\sim 1$ because of the $t$-channel
electron exchange amplitude, Fig. 1(b).
\par
The fraction of the transversely polarized $J/\psi$ is displayed in Fig. 3(c) 
and (d) where the transverse cross sections have been normalized by the 
respective total cross sections for the three production modes. For 
the $J/\psi c{\bar c}$ mode, both the QCD and QED cross sections are dominated
by the transverse $J/\psi$, and the transverse $J/\psi$ fraction  
increases with the $J/\psi$ momentum. On the other hand,
an opposite behavior is obtained for $J/\psi gg$. 
We point out here that the transverse fraction for the QCD $J/\psi c {\bar c}$
production approaches to unity as $z$ reaches 0.90, the maximum for the 
process. This feature is not clearly seen in Fig. 3(c) due to
the finite bin size, but has been checked numerically. 
It indicates that all $J/\psi$ mesons are transversely polarized
if the associated $c {\bar c}$ has vanishing relative momentum.
This is consistent with the exclusive double-charmonium production
$e^+ e^- \to J/\psi \eta_c$ where all $J/\psi$ mesons are transversely 
polarized. The scattering-angle distributions are also quite distinct from
each other; as shown in Fig. 3(d). In the region of the small scattering 
angles, the transverse components for the QCD $J/\psi c {\bar c}$ and 
$J/\psi gg$ processes are comparatively larger than those in the large-angle 
region. For the QED $J/\psi c {\bar c}$ process, the transverse fraction rapidly
falls off in the region $|\cos\theta|>0.8$.
\begin{figure}
\centerline{\epsfysize 5 truein \epsfbox{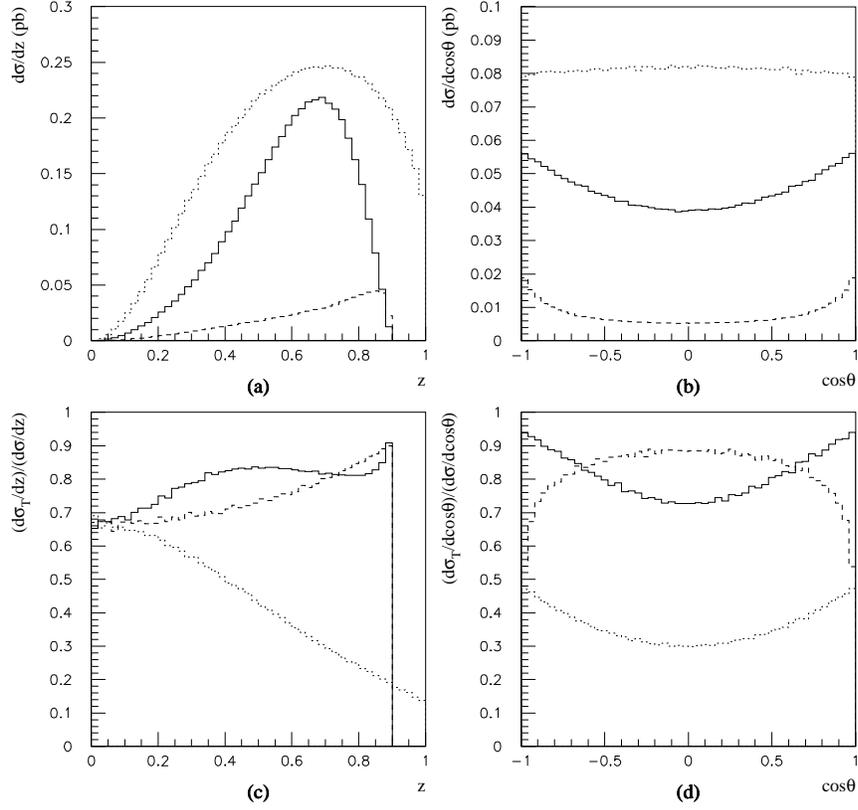}}
\caption{
Momentum and angular distributions of $e^+ e^-\to J/\psi c{\bar c}$ and 
$J/\psi gg$ processes:
(a) and (b) are for the total cross section, and (c) and (d) are for the 
fraction of the transversely polarized $J/\psi$ production.
The solid, dashed  and dotted lines correspond to 
the QCD $J/\psi c {\bar c}$, QED $J/\psi c {\bar c}$,
and $J/\psi gg$ production, respectively. }
\end{figure}

\section{The Collinear Suppression and $K$ Factors}  
\par
We note that although the contribution from the $J/\psi gg$ process 
is significant in the end-point region $0.90<z<1$ in our leading-order 
calculations, there is no such apparent signal observed by experiments. 
The probable reason of this discrepancy is that NRQCD does not include 
collinear degrees of freedom at large $z$ and therefore both the perturbative 
expansion and the operator product expansion (OPE) break down~\cite{pe-ope}. 
The problem can be cured by combining NRQCD for the heavy degrees of freedom 
with the soft-collinear effective theory (SCET) for the light degrees of 
freedom~\cite{scet}. A similar problem occurs in the radiative decay 
$\Upsilon \to \gamma gg$ at large $\gamma$ energies as mentioned in Sec. 1. 
In Refs.~\cite{fleming}, it has been shown that the prediction of SCET 
in the end-point region of the $\gamma$ energy distribution 
is much closer to the CLEO data than the leading-order calculations in NRQCD.
\par
In this paper instead of performing the complicated calculations within SCET, 
we adopt a phenomenological approach to obtain an appropriate end-point 
spectrum for the $J/\psi gg$ process by noting that in SCET, the operator of
$\gamma^* \to J/\psi g g$ is
in the same form as that of $\Upsilon \to \gamma g g$. It is further observed
that, the collinear Sudakov factor of $\Upsilon$ decay arises solely from the
gluon jet function. In the leading power expansion of SCET,
the jet function depends only on the large light-cone momentum component. It is
easy to find that, the large momentum component of two-gluon jet of
$\Upsilon \to \gamma g g$ decay ($M_\Upsilon$) is roughly equal to that of
$\gamma^* \to J/\psi g g $ process ( $(s-M_{J/\psi}^2)/ \sqrt{s}$).
This implies that the collinear Sudakov factor of these two processes are quite
similar. Therefore it should be a reasonable approximation that these two 
processes have the same collinear suppression function.
\par 
Accordingly, we disentangle the collinear 
suppression function $F(z')$ from the experimental data and the tree level
prediction for $\Upsilon \to \gamma gg$ as
\footnote{We obtain the experimental data and the tree level prediction
from Fig. 3 in the first reference in Refs.~\cite{fleming}.}
\bea
\frac{d\Gamma_{exp}}{dz'}=F(z')\times \frac{d\Gamma_{LO}}{dz'},
\eea
where $\Gamma_{exp}$ and $\Gamma_{LO}$ are the decay width for
the experimental measurement and the leading-order calculation respectively. 
$z'$ is the normalized photon energy and can be expressed
as a function of the invariant mass of the two gluons
\bea 
z'=\frac{2 E_\gamma}{M_\Upsilon}=1-\frac{M_{gg}^2}{M_\Upsilon^2}.
\eea
We assume that the function $F(z')$ can be parameterized as
\bea \label{func}
F(z')=
\left\{
\begin{array}{ll}
1&for~0\leq z'\leq 0.5 \\
(1-z'){\rm exp}(c_1 z'+c_2 {z'}^2)&    for~0.5<z'\leq1
\end{array}
\right.
\eea
with two free parameters $c_1$ and $c_2$. We find a good fit to the data and 
by using the LO calculation presented in Ref.~\cite{fleming}, for
\bea\label{c1c2}
c_1=0.96,~~~c_2=0.69.
\eea
The suppression function in Eq. (\ref{func}) shows
that the collinear effects in the region $0\leq z'\leq 0.5$ are
negligible, the large suppression takes place in the end-point region
$0.7 < z' \leq 1$. 
With Eqs. (\ref{func}) and (\ref{c1c2}), the suppression effects in the
$\Upsilon$ radiative decay can be well described.
Finally, we obtain the modified differential cross section for
$e^+ e^- \to J/\psi gg$
\bea
\frac{d\sigma_{gg}}{z}=F\left(\sqrt{z^2+\frac{4M_{J/\psi}^2}{s}}\right)
\times \frac{d\sigma_{gg}^{LO}}{dz}, 
\eea
where we replace the photon energy fraction $z'$ by the $J/\psi$ energy
fraction $2E_{J/\psi}/\sqrt{s}=\sqrt{z^2+4M_{J/\psi}^2/s}$. We expect
that this is a good approximation because the energy scale of the two processes 
is roughly the same.
Integrating over $z$, the total cross section is reduced to
\bea\label{sup}
\sigma_{gg}&=&0.123~{\rm pb,}
\eea
which is about 24\% smaller than the original value in Eq. (\ref{lead}).
\par
The end-point suppression behavior is plotted in Fig. 4(a) by the solid line,
while the result of the leading order calculation is shown by the dashed line
for comparison. In terms of the momentum fraction variable $z$, the suppression 
effect starts at around $z=0.4$. As $z$ is close to unity, the
differential cross section drops to zero, which diminishes the disagreement
with the Belle data. The angular distribution is modified slightly,
as shown in Fig. 4(b). The cross section in the central region ($\cos\theta
\sim 0$) is suppressed most because that the large $z$ component dominated by
the longitudinally polarized $J/\psi$ mesons is suppressed significantly. 
After the collinear suppression factor is introduced, there is no
apparent disagreement between the Belle data at large $z$ and the NRQCD 
production of the $J/\psi gg$ process, without introducing a further overall
suppression factor. 
\begin{figure}
\centerline{\epsfysize 2.5 truein \epsfbox{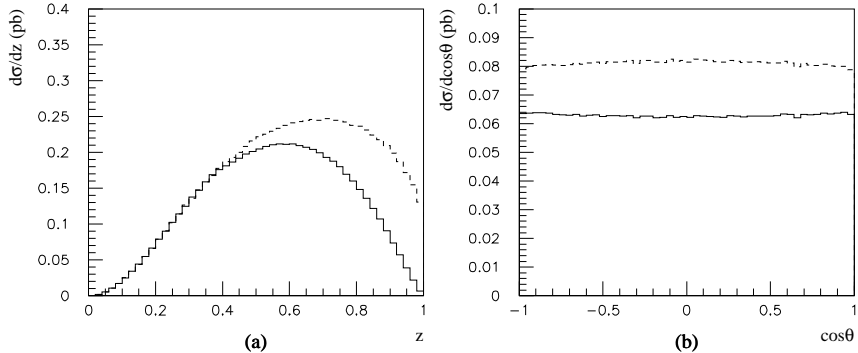}}
\caption{
Momentum (a) and angular (b) distributions for the $J/\psi gg$ cross section.
The dashed line is for the leading order calculation, while
the solid line is for the calculation with the collinear suppression.
}
\end{figure}
\par
The NRQCD prediction for the direct $J/\psi c {\bar c}$ productions
is however significantly smaller than the Belle data. The prompt 
$J/\psi c {\bar c}$ production cross section obtained by Belle is 
$\sigma_{prompt}=(0.87^{+0.21}_{-0.19}\pm 0.17)$ pb based on 
$D^{*+}$ and $D^0$ measurements~\cite{belle2}. 
Because the prompt $J/\psi$ data contain contributions from 
the $\psi(2S)\to J/\psi$ transitions, we estimate the direct $J/\psi c{\bar c}$
production cross section as follows. Working in the framework of NRQCD, 
we find that direct $\psi(2S)c {\bar c}$ and $J/\psi c {\bar c}$ cross sections
are proportional to the meson wave functions at the origin, and to the leptonic
decay widths of $\psi(2S)$ and $J/\psi$. According to Eq. (\ref{wave}), we have
\bea
\sigma_{dir}(\psi(2S)c{\bar c}):\sigma_{dir}(J/\psi c {\bar c})
= M_{\psi(2S)}^2\Gamma(\psi(2S)\to e^+e^-): M_{J/\psi}^2
\Gamma(J/\psi\to e^+e^-).
\eea
By using $\Gamma(\psi(2S)\to e^+e^-)= 2.15\times 10^{-6}\gev$ and the branching
ratio for the $\psi(2S)\to J/\psi X$ transition fraction $B=55.7\%$~\cite{pdg},
we obtain the direct $J/\psi c{\bar c}$ cross section
\bea\label{dir}
\sigma_{dir}(J/\psi c {\bar c})&=&\sigma_{prompt}\left[1+
B(\psi(2S)\to J/\psi X)\times 
\frac{M_{\psi(2S)}^2 \Gamma(\psi(2S)\to e^+e^-)}
{M_{J/\psi}^2 \Gamma(J/\psi\to e^+e^-)}\right]^{-1} \nonumber \\
&\simeq& 0.66\pm 0.26~{\rm pb}.
\eea
Here we estimate the error of $\sigma_{dir}$ from the statistical
and systematic errors of $\sigma_{prompt}$. Comparing with the result given 
in Eq.~(\ref{lead}), 
$\sigma_{cc}(J/\psi c{\bar c}) =\sigma_{cc}^{QCD}+\sigma_{cc}^{QED}=0.11$ pb,
there is still a large gap between the experimental central value
and the theoretical prediction. Moreover, the rate of $\sigma_{cc}/(\sigma_{cc}+
\sigma_{gg})$ is about 0.46 evaluated from Eqs. (\ref{lead}) and (\ref{sup}), 
much smaller than the current experimental measurement
$0.67\pm 0.12$ in Eq.~(\ref{rate2}) as indicated in Sec. 1.

According to the argument given in the first section, the possible source 
of the disagreement may arise from the large higher-order corrections 
to the QCD $J/\psi c {\bar c}$ production process.  For instance, 
if we set as $K$ factor to unity for both QED $J/\psi c {\bar c}$ and 
$J/\psi gg$ processes,
the direct $J/\psi c {\bar c}$ cross section in Eq. (\ref{dir}) requires a $K$ 
factor for the QCD $J/\psi c {\bar c}$ production to be
\bea\label{k1}
K=7.2\pm 2.9,
\eea
while the fraction of the $J/\psi c {\bar c}$ process in Eq. (\ref{rate2})
gives rise to 
\bea\label{k2}
K=2.6^{+2.4}_{-1.1}.
\eea
To fit to both the cross section and the fraction, we set the $K$ factor 
to a moderate value 4 for the QCD $J/\psi c {\bar c}$ process,
and then find
\bea
\sigma_{cc}=K \sigma_{cc}^{QCD}+\sigma_{cc}^{QED}=0.374~{\rm pb},
~~~\sigma_{cc}/(\sigma_{cc}+ \sigma_{gg})=75\%.
\eea
A part of the large $K$ factor may come from the uncertainty in the wave 
function at the origin. By replacing the leading-order formula of the 
leptonic $J/\psi$ decay in Eq. (\ref{wave}) by the next-to-leading-order one, 
the value of the wave function $|R(0)|^2$ is changed to be 0.792 GeV$^3$
~\cite{braaten-lee}, while the wave function from some potential models 
is even larger, such as 0.810 GeV$^3$~\cite{quigg}. 
Because the cross section is proportional to $|R(0)|^2$, this effect 
gives the $K$ factor of $792/447\sim 1.77$ and $810/447\sim 1.81$ respectively.
Another factor of two from the higher-order corrections to the hard scattering
part may lead to the combined effect as large as $K\sim 4$.
For the QED $J/\psi c{\bar c}$ case, since the fragmentation configurations 
are dominant, the ratio $\sigma_{cc}^{QED}/\Gamma(J/\psi\to e^+ e^-)$ 
is approximately a constant to any order of QCD perturbation theory. 
For this reason we choose the corresponding 
$K$ factor to be equal to one. The $J/\psi gg$ cross section can also be 
enlarged by a factor of two caused by the the next-to-leading-order value 
of $|R(0)|^2$, but the higher-order QCD corrections
of the hard part is unknown. For simplification and consistency with 
experimental data we set its $K$ factor to be about one.

We emphasize here that only the qualitative aspects of the above discussions
on the $K$ factors should be taken seriously. The $K$ factors should be 
calculated explicitly, and 
the predictions should be confronted against experiments. In the next section
we show how to distinguish the contributing subprocesses experimentally.

\section{Spin Density Matrix and Sensitivity to $K$ Factors}
\par
In this section, we give more detailed analyses on the elements of
the spin density matrix and various distributions. 
\par
Since the $J/\psi$ mesons are reconstructed experimentally by using the 
leptonic decays $J/\psi \to \mu^+\mu^-,e^+ e^-$, we describe the decay 
processes in the $J/\psi$ rest frame by the polar and the azimuthal angle,
$\theta^*$ and $\phi^*$. The polar axis is chosen along the $J/\psi$ 
momentum in the $e^+ e^-$ c.m. frame. The reduced helicity amplitudes 
for $e^+ e^- \to J/\psi X$ are denoted by $M_\lambda$, where 
$\lambda=+1,0,-1$ is the $J/\psi$ helicity. Other particle's
helicities are unobserved (integrated out) here. Then the production density 
matrix can be expressed as $R_{\lambda\lambda'}=\Sigma M_\lambda 
M_{\lambda'}^*$, where the summation is over the other particle helicities
and momenta. For $e^+e^-\to J/\psi X\to l^+ l^- X$, the differential cross 
section can be expressed as
\bea\label{density}
& &\frac{d\sigma_{total}}{dz d\cos\theta d\cos\theta^*d\phi^*} \nonumber\\
&=&\frac{3}{4\pi}\left\{ (R_{++}+R_{--})\frac{1}{4}(1+\cos^2\theta^*)
+R_{00}\frac{1}{2}(1-\cos^2\theta^*)\right. \nonumber\\
&+&2\re(R_{+0}-R_{0-})\frac{1}{4\sqrt{2}}\sin 2\theta^*\cos\phi^*
-2\im(R_{+0}-R_{0-})\frac{1}{4\sqrt{2}}\sin 2\theta^*\sin\phi^* \nonumber\\
&+&\left. 2\re(R_{+-})\frac{1}{4}\sin^2\theta^*\cos 2\phi^*
-2\im(R_{+-})\frac{1}{4}\sin^2\theta^*\sin 2\phi^*\right\}.
\eea
Here $R_{\lambda\lambda'}$ are the functions of $z$ and $\cos\theta$ and we use
the normalization $d\sigma_{total}/dz d\cos\theta=(R_{++}+R_{--}+R_{00})$
after integrating over $\cos\theta^*$ and $\phi^*$. From Eq. (\ref{density}), 
one can find that there are six independent combinations 
of the density matrix elements for the $J/\psi$ productions that can be
measured through the decay-lepton angular distributions.
\par
According to the above common form, we calculate the different density matrix
elements for all $J/\psi$ inclusive processes and find that two of them
$\im(R_{+0}-R_{0-})$ and $\im(R_{+-})$ vanish after integration over the
unobserved internal momenta of $c{\bar c}$ or $gg$ system because of CP
invariance. Among the surviving four terms,
$(R_{++}+R_{--})$ is actually the cross section for
the transversely polarized $J/\psi$ and $R_{00}$ is for the 
longitudinally polarized $J/\psi$. After integrating over $z$ and $\cos\theta$,
three of them survive
\bea\label{recom}
\sigma_L&=&\int R_{00} dz d\cos\theta, \nonumber \\
\sigma_T&=&\int (R_{++}+R_{--}) dz d\cos\theta, \nonumber \\
\sigma_{+-}&=&\int 2{\re(R_{+-})} dz d\cos\theta. 
\eea
where $\sigma=\sigma_L+\sigma_T$ is the total cross section.  We also note 
that $\sigma_{+-}$ is from the nondiagonal spin density matrix elements
and should be measured through the azimuthal angle of the decay leptons.
The term proportional to $\re(R_{+0}-R_{0-})$ vanished after integration over
$\cos\theta$.
\par
We plot the momentum and angular distributions for $\sigma$, $\sigma_T$
and $\sigma_{+-}$ respectively in Fig. 5. The solid lines are for the sum
of QCD and QED $e^+ e^- \to J/\psi c {\bar c}$ processes, where we set 
$K=4$ factor for the QCD process. The dashed lines represent the 
$e^+ e^-\to J/\psi gg$ contribution where the collinear suppression factor 
has been included. The dotted lines show their sum, the total $J/\psi$ 
inclusive cross sections. Due to its large $K$ factor, the 
$J/\psi c {\bar c}$ contribution is now dominant. Therefore, as shown in 
Fig. 5(a), our prediction of the total differential 
cross section versus the $J/\psi$ momentum is roughly consistent with the 
experimental observation. The small but finite contribution from the 
$J/\psi gg$ process at $z>0.9$ should eventually be observed by experiments,
and its magnitude will constrain the $K$ factor for the $J/\psi gg$ process
for which we set $K=1$ in our calculation.
From Fig. 5(b), we see that the $J/\psi c{\bar c}$ mode prefers large 
$|\cos\theta|$, while the $J/\psi gg$ mode gives rather flat $\cos\theta$
distribution. Comparing with the transverse fractions for the QCD 
$J/\psi c {\bar c}$ process displayed in Fig. 3(c) and (d), we find those for 
the summed $J/\psi c {\bar c}$ mode in Fig. 5(c) and (d) are changed only
slightly because of the large $K$ factor for the QCD process. In the region 
$|\cos\theta|>0.9$, the small suppression from the QED $J/\psi c {\bar c}$ 
component is observed. The nondiagonal element $\sigma_{+-}$ is quite different
for $J/\psi c {\bar c}$ and $J/\psi gg$, in particular at large $z$ where 
$\sigma_{+-}$ has the opposite signs. Since $\sigma_{+-}$ is the coefficient of
$\cos2\phi^*$ depicted in Eq. (\ref{density}), it should approach zero as the 
direction of the $J/\psi$ momentum is along the beamline, \ie $|\cos\theta|=1$.
It can be seen from the angular distributions for $\sigma_{+-}$ plotted in 
Fig. 5(f).
\begin{figure}
\centerline{\epsfysize 7 truein \epsfbox{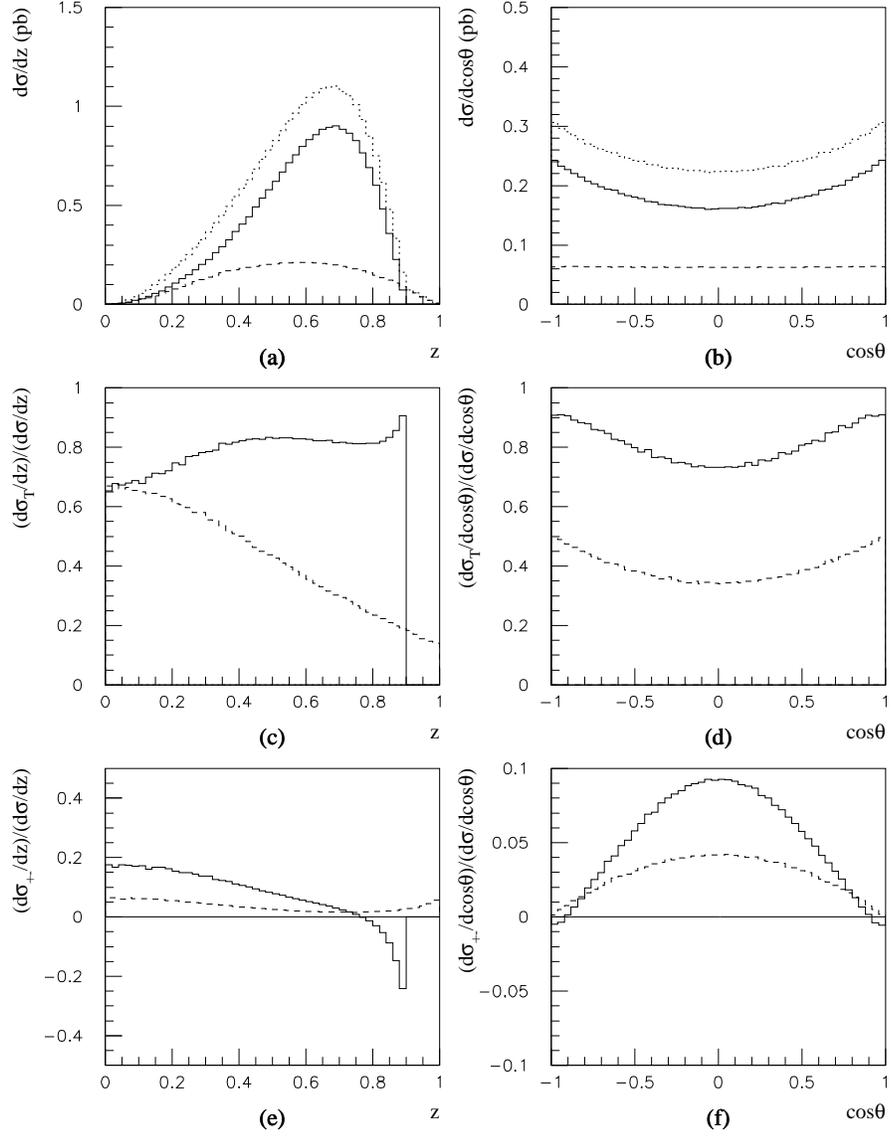}}
\caption{
Momentum and angular distributions:
(a) and (b) are for the total cross section, and (c) and (d) are for the 
the fraction of the transversely polarized $J/\psi$,
(e) and (f) are for the fraction of $\sigma_{+-}$.
The solid, dashed  and dotted lines correspond to 
the $J/\psi c {\bar c}$, $J/\psi gg$ and total inclusive productions
respectively. }
\end{figure}
\par 
In Fig. 6, we extend our above analyses on the angular distributions for
different momentum regions: $0\leq z \leq 0.4$(1-a,b,c),
$0.4 < z \leq 0.7$(2-a,b,c) and $0.7< z \leq 1$(3-a,b,c), corresponding to
$0\leq P_{J/\psi} \leq 1.94\gev$, $1.94\gev< P_{J/\psi} \leq 3.40\gev$ and
$3.40\gev< P_{J/\psi} \leq 4.86\gev$, respectively. The angular distribution
of the total cross section $d\sigma/d\cos\theta$ is given in the figures
(1-a,2-a,3-a), where the $J/\psi gg$ contribution is almost flat at all $z$
region, and the contribution from the $J/\psi c{\bar c}$ process
is more pronounced at large
$|\cos\theta|$. The ratio of $d\sigma/d\cos\theta$ at $|\cos\theta|=1$
and $\cos\theta=0$ is about 1.3 at small $z$(1-a), 1.5 at medium $z$(2-a),
and 1.7 at large $z$(3-a) for the $J/\psi c{\bar c}$ contribution.
The fractions of the transverse cross section $d\sigma_T/d\cos\theta$ and the
total distribution are shown in the figures (1-b,2-b,3-b). 
%
The $J/\psi c{\bar c}$ contribution has generally large $\sigma_T/\sigma$ ratio,
greater than about 0.7, at all $\cos\theta$ and $z$ regions, while
the $J/\psi gg$ contribution gives smaller ratio, especially at large angles
and large $z$. It should also be noted that the $\cos\theta$ dependence of the
$\sigma_T/\sigma$ ratio for $J/\psi c{\bar c}$ process is sensitive to the
$K$ factor for the QCD process, because the QCD and QED subprocesses have
very different $\cos\theta$ distribution, as shown in Fig. 3(d).
The fractions of $\sigma_{+-}/d\cos\theta$ are displayed in the figures 
(1-c,2-c,3-c). Similar with
the distributions in Fig. 5(f), the fraction of $\sigma_{+-}/d\cos\theta$ is
zero at $|\cos\theta|=1$. For the $J/\psi gg$ contribution, the maximum
value of the fraction of $\sigma_{+-}/d\cos\theta$ appears at
$\cos\theta=0$ and is about 0.08 at small $z$(1-c), 0.04 at medium $z$(2-c),
and 0.03 at large $z$(3-c). For the $J/\psi c{\bar c}$ contribution,
the fraction of $\sigma_{+-}/d\cos\theta$ at $\cos\theta=0$ is about 0.22, 0.12
and -0.02 at small $z$(1-c), medium $z$(2-c) and large $z$(3-c) respectively.
At large $z$(3-c), the fraction of $\sigma_{+-}/d\cos\theta$ is negative due to
the large contributions from the QED $J/\psi c{\bar c}$ process. Summing up, the
angular distributions for the $J/\psi c{\bar c}$ and $J/\psi gg$ are 
significantly different in the various momentum regions.
\begin{figure}
\centerline{\epsfysize 6 truein \epsfbox{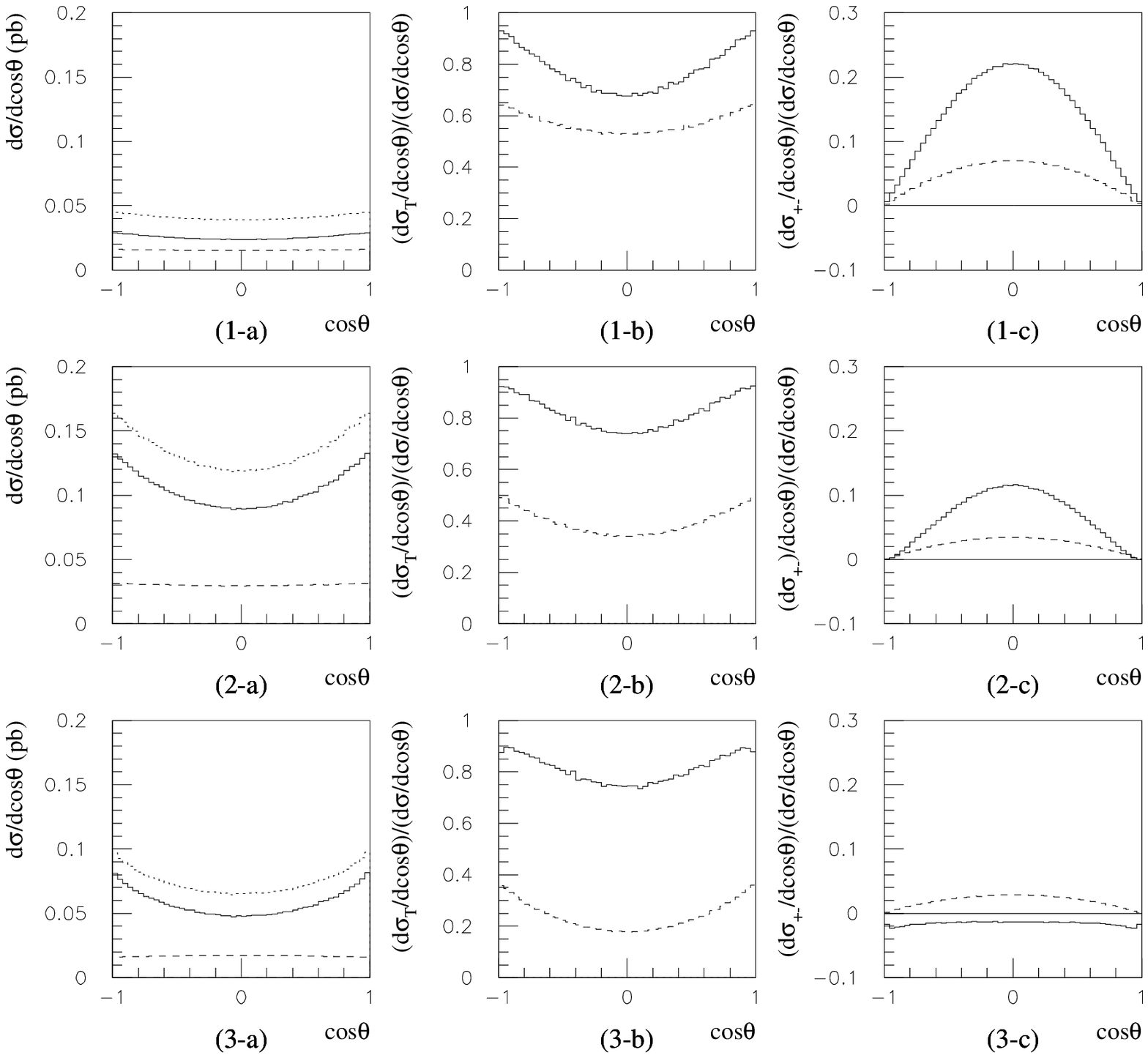}}
\caption{
Angular distributions: (1-a,b,c), (2-a,b,c) and (3-a,b,c) are plotted in the
intervals $0\leq z \leq 0.4$, $0.4 < z \leq 0.7$ and $0.7< z \leq 1$ respectively.
(1,2,3-a) are for the total cross section, (1,2,3-b) are for the fraction of
the transversely polarized $J/\psi$ and (1,2,3-c) are for the fraction of
$\sigma_{+-}$.  The solid, dashed and dotted lines correspond to
the $J/\psi c {\bar c}$, $J/\psi gg$ and total inclusive productions
respectively. }
\end{figure}
\par
The results of Fig. 5 and Fig. 6 are shown for $K=4$ for the QCD
$J/\psi c {\bar c}$ production. Since the actual value of $K$ is
still uncertain ($K=7.2\pm 2.9$ from $\sigma_{dir}(J/\psi c {\bar c})$,
and $K=2.6^{+2.4}_{-1.1}$ from $\sigma(J/\psi c {\bar c})/\sigma(J/\psi X)$
as shown in Eqs. (\ref{k1}) and (\ref{k2})),
we study the sensitivity of various distributions on $K$.
In order to give a more
direct comparison to the experimental data, we introduce a few
kinematic asymmetries. The first one is the observable $\alpha$
defined in Eq. (\ref{cos}), which can be obtained from
the $J/\psi$ scattering angle distributions.
Although the angular distribution for the QED $J/\psi c{\bar c}$
production is not a standard $1+\alpha\cos^2\theta$ function, for
comparing with experimental data, we can still use
an effective function just like Eq. (\ref{cos}) to fit the curves
for the QED mode in the central region, $|\cos\theta<0.9|$,
and then obtain an effective $S_{cc}^{QED}$ and
$\alpha_{cc}^{QED}$\footnote{Here the fitting program PAW (Physics
Analysis Workstation) and $\chi^2$ method have been utilized to
obtain the $\alpha$ values.}. The angular distribution for the QED
mode
is described effectively by
\bea\label{eff-cos}
\frac{d^2 \sigma_{cc}^{QED}}{dz d\cos\theta}
\simeq S_{cc}^{QED}(z)[1+\alpha_{cc}^{QED}(z)\cos^2\theta].
\eea
Due to the small cross section of the QED $J/\psi c{\bar c}$ and according to
the traditional experimental data analysis, the technique is sufficient
to give a good accuracy in comparison with the experimental data.
\par
Simply summing over Eqs. (\ref{cos}) and (\ref{eff-cos}),
we work out the effective angular coefficients for
$J/\psi c{\bar c}$ production and the total inclusive process respectively,
\bea
\alpha_{cc}=\frac{S_{cc}^{QED}\alpha_{cc}^{QED}+S_{cc}^{QCD}\alpha_{cc}^{QCD}}
{S_{cc}^{QED}+S_{cc}^{QCD}},
~~~
\alpha=\frac{S_{cc}^{QED}\alpha_{cc}^{QED}+S_{cc}^{QCD}\alpha_{cc}^{QCD}
+S_{gg}\alpha_{gg}} {S_{cc}^{QED}+S_{cc}^{QCD}+S_{gg}}.
\eea
\par
The second observable is relative to the polar angle $\theta^*$ in $J/\psi$
leptonic decays. One can integrate over all other parameters but leave only $\theta^*$
in Eq. (\ref{density}), then obtain an reduced formula
\bea\label{cos-star}
\frac{d\sigma}{d\cos\theta^*}=S^*[1+\alpha^*\cos^2\theta^*],
\eea
which is analogue to Eq. (\ref{cos}).
The coefficient $\alpha^*$ has a simple relation with the total cross
section $\sigma$ and the transverse cross section $\sigma_T$:
\bea
\alpha^*=\frac{-2\sigma+3\sigma_T}{2\sigma-\sigma_T}.
\eea
\par
The observables $\alpha$ and $\alpha^*$ versus the $K$ factor for the QCD
$J/\psi c{\bar c}$ process are displayed in Fig. 7 in the full momentum
spectrum $0\leq z \leq 1$(1-a,1-b) as well as in the
moderate and large $z$ regions $0.4 < z \leq 0.7$(2-a,2-b) and
$0.7< z \leq 1$(3-a,3-b) respectively.
The integrated luminosity in the region $0\leq z\leq 0.4$ at the B factories
comes from the off-resonance data and hence is one order smaller than
those in the moderate and large $z$ regions. It is hard to measure the
observables precisely in $0\leq z\leq 0.4$, so we do not present them here.
The dashed, dotted and solid lines correspond to the $J/\psi c {\bar c}$,
$J/\psi gg$ and the total inclusive productions respectively.
The $\alpha$ versus $K$ is given in figures (1-a,2-a,3-a).
The $\alpha$ parameter for the $J/\psi gg$ contribution 
(independent of the $K$ factor)
is almost zero at all $z$ region, which is consistent with the flat curves
for the total distribution cross section in Figs. 5(b) and 6(1-a,2-a,3-a).
For the $J/\psi c{\bar c}$ contribution, $\alpha$ is sensitive to the $K$
values, in particular at small $K$ and at large $z$(3-a), it varies from
0.9 to 0.6 for $1<K<5$. This is due to the large QED $J/\psi c{\bar c}$
contribution in large $z$ region. The $\alpha$ parameter for the total inclusive
process varies from 0.25 to 0.4 for $0\leq z \leq 1$(1-a), from 0.3 to 0.42 for
$0.4 < z \leq 0.7$(2-a), and from 0.32 to 0.52 for $0.7< z \leq 1$(3-a).
$\alpha^*$ versus $K$ is shown in figures (1-a,2-a,3-a).
The $\alpha^*$ parameter for the $J/\psi c{\bar c}$ contribution is about 
0.4 at all $z$ region almost independent of $K$,
while for the $J/\psi gg$ contribution it is about $-$0.5 for
$0.4<z<0.7$ (2-b), $-$0.7 for $0.7<z<1$ (3-b).
For the total inclusive $J/\psi$ process, the $\alpha^*$ is more
sensitive to the $K$ factor than the $\alpha$. It varies from -0.82 to 0.24 for
$0\leq z \leq 1$(1-b) and $0.4 < z \leq 0.7$(2-b), and from -0.7 to 0.2
for $0.7< z \leq 1$(3-b).
We also plot the Belle data with a one-sigma error which fall into
the hatched regions. As shown in Fig. 7, our theoretical calculations do not
satisfy the current data very well. The data for $\alpha$
are close to the case of $J/\psi c {\bar c}$ only, while for $\alpha^*$
there is a tendency towards the $J/\psi gg$ predictions.
However the errors of the experimental data are still large,
and more accurate measurements are needed.
\begin{figure}
\centerline{\epsfysize 7 truein \epsfbox{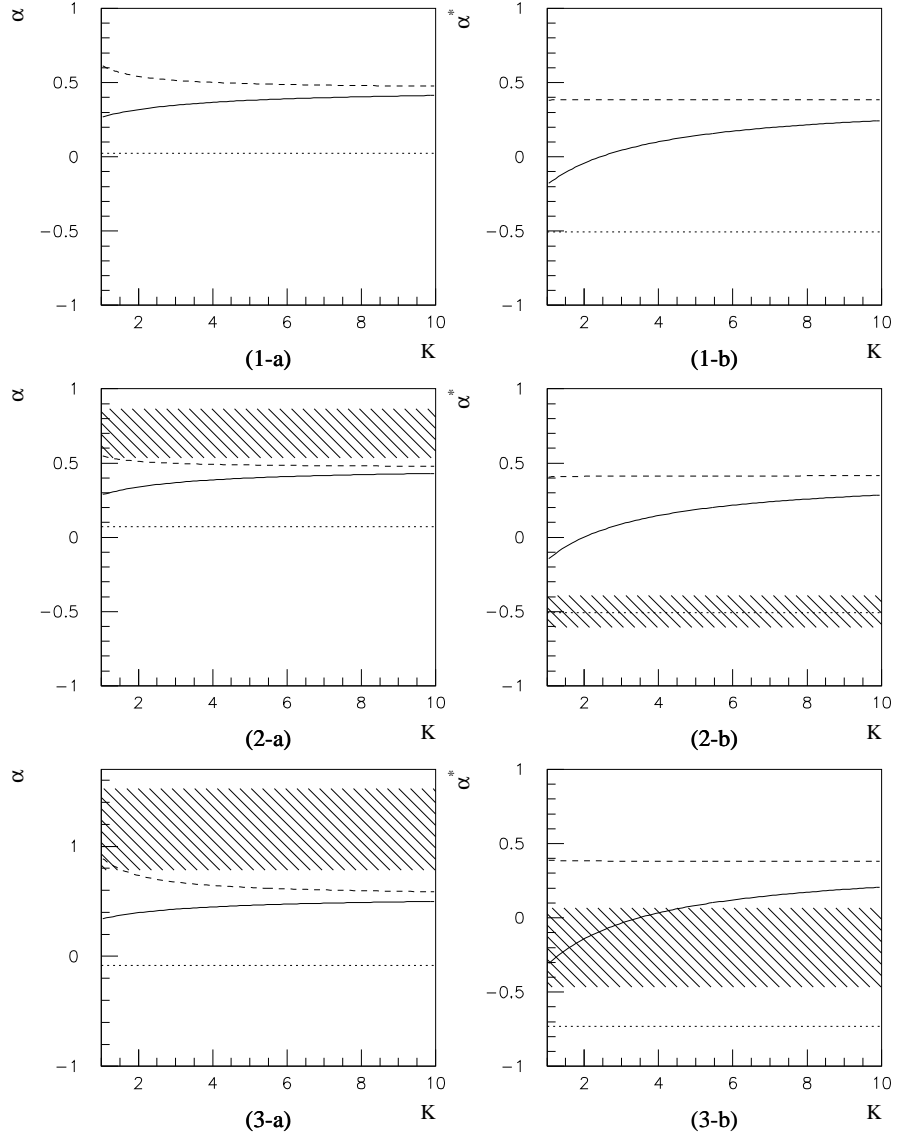}}
\caption{
Sensitivity of the $K$ factor for $\alpha$ and $\alpha^*$:
(1-a,b), (2-a,b) and (3-a,b) are plotted in the
intervals $0\leq z \leq 1$, $0.4 < z \leq 0.7$ and $0.7< z \leq 1$ respectively.
(1,2,3-a) and (1,2,3-b) are for various asymmetries $\alpha$ and
$\alpha^*$ respectively. The dashed, dotted and solid lines correspond to
the $J/\psi c {\bar c}$, $J/\psi gg$ and the total inclusive productions respectively.
The experimental data with a one-sigma error fall into the hatched regions.
}
\end{figure}
%
\par
The third asymmetry can be obtained by integrating over all other parameters but
leave only the azimuthal $\phi^*$ in Eq. (\ref{density}) and therefore the
off-diagonal density-matrix element $\sigma_{+-}$ can be measured. From Eq.
(\ref{density}), we find that $\sigma_{+-}$ is proportional to $\sin^2\theta^*$.
We also notice from Fig. 6 that $\sigma_{+-}$ is much larger in the small
$|\cos\theta|$ region.
In order to enhance the signal as large as possible,
we impose the angular cuts $|\cos\theta|,|\cos\theta^*|\leq 1/4$, obtaining
\bea\label{phi-star}
\frac{d\sigma_{total}}{d\phi^*}\left(|\cos\theta|,|\cos\theta^*|\leq \frac{1}{4}\right)
=\frac{1}{512\pi}
\left(94\sigma-45\sigma_T+47\sigma_{+-}\cos2\phi^*\right).
\eea
To retain the information of $\phi^*$,
we split the phase space in terms of $\phi^*$ into eight parts and
integrate over them separately. After recombination of the eight integrations
and normalization by the total cross section, the asymmetry is constructed as
\bea
A\left(|\cos\theta|,|\cos\theta^*|\leq \frac{1}{4}\right)
&=&\frac{\sum_{n=0}^3\left[\int_{n\pi/2}^{(n/2+1/4)\pi}d\phi^*
-\int_{(n/2+1/4)\pi}^{(n+1)\pi/2}d\phi^*\right]
\times({d\sigma_{total}}/{d\phi^*})}
{\sum_{n=0}^3\left[\int_{n\pi/2}^{(n/2+1/4)\pi}d\phi^*
+\int_{(n/2+1/4)\pi}^{(n+1)\pi/2}d\phi^*\right]
\times({d\sigma_{total}}/{d\phi^*})} \nonumber \\
&=&\frac{94\sigma_{+-}}{94\pi\sigma-45\pi\sigma_T}.
\eea
\par
The asymmetry $A$ is generally small except at very high $z$, because
$|\sigma_{+-}|$ for the $J/\psi c{\bar c}$ production is quite large
near $z\sim 0.9$ which is shown in Fig. 5(e). In Fig. 8, we plot the 
sensitivity of the $K$ factor for the asymmetry $A$ in the
interval $0.85 \leq z\leq 1$. $A$ for the $J/\psi gg$ process (the dotted line)
is almost zero while for the $J/\psi c {\bar c}$ (the dashed line) can reach
-0.11. Therefore $A$ for the total inclusive production (the solid line) is
sensitive to the $K$ factor due to the enhancement of the $J/\psi c {\bar c}$
component with the increasing $K$ value.
\begin{figure}
\centerline{\epsfysize 3 truein \epsfbox{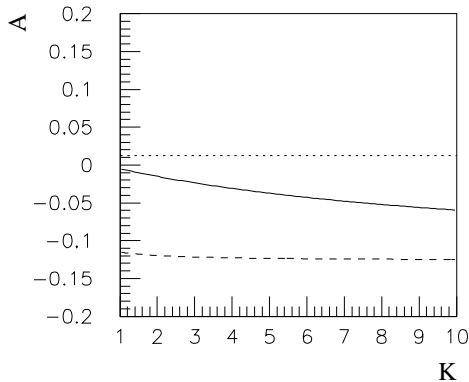}}
\caption{
Sensitivity of the $K$ factor for the asymmetry $A$ in the interval $0.85
\leq z\leq 1$ . The dashed, dotted and solid lines correspond to
the $J/\psi c {\bar c}$, $J/\psi gg$ and the total inclusive productions respectively.
}
\end{figure}
\section{Relation between the Inclusive and Exclusive Processes}
\par
Since the introduction of the large $K$ factor for the QCD
$J/\psi \ccbar$ process could provide a reasonable account of
the inclusive $J/\psi$ production data, especially the total
rate and the $J/\psi\ccbar$ fraction, it may also give us a
hint of the solution to the puzzle of exclusive $J/\psi\eta_c$
production, where the observed cross section exceeds the
NRQCD prediction by one order of magnitude. For instance,
the $J/\psi\eta_c$ production cross section can be enhanced by
a squared of the $K$ factor for the $J/\psi\ccbar$ production
process.  This can happen if the hard part receives large
higher order corrections only when $\ccbar$ pair is parallel.
Numerically, if we take $K=4$ for the inclusive production,
the $K$ factor for the exclusive one can be as large as 16,
which is consistent with the estimate given in Ref.~\cite{hagiwara-kou-qiao}.
%
%
\par
In the case that $J/\psi$ mesons take maximum momenta in 
$e^+ e^-\to J/\psi c{\bar c}$ where $c {\bar c}$ pair have the same momentum,
the inclusive process matches the exclusive process. Therefore we expect that 
near the high end point of the $J/\psi$ momentum, the inclusive process 
should have similar angular distributions with the exclusive one.
Notice that Eq.~(\ref{density}) is generally valid for both inclusive
and exclusive $J/\psi$ processes.
Similar to what has been done in the last section,
the angular distribution for $e^+ e^-\to J/\psi \eta_c$ can be expressed as
a more definite form ~\cite{hagiwara-kou-qiao},
\bea\label{density2}
\frac{d\sigma}{d\cos\theta d\cos\theta^*d\phi^*}\sim
(1+\cos^2\theta)(1+\cos^2\theta^*)-\sin^2\theta\sin^2\theta^*\cos 2\phi^*,
\eea
where the definitions of $\theta$, $\theta^*$ and $\phi^*$ are the same
as in Eq.~(\ref{density}). Comparing the above equation with 
Eq.~(\ref{density}), we get some general information:
\begin{itemize}
\item It is well known that there is no longitudinally polarized $J/\psi$
produced in the exclusive $J/\psi \eta_c$ process due to parity conservation; 
$\sigma_L/\sigma \to 0$ in Eq.~(\ref{recom}).
The feature is also reflected in Eq. (\ref{density2}).
\item After integrating over $\theta^*$ and $\phi^*$ in Eq. (\ref{density2}),
the total (transverse) cross section $\sigma$ is proportional to 
$(1+\cos^2\theta)$. It means that the coefficient $\alpha$ defined in 
Eq. (\ref{cos}) is equal to one in the exclusive limit.
\item The nondiagonal density-matrix element $\sigma_{+-}$ is proportional
to $\sin^2\theta$.
\item If the scattering angle $\theta$ is fixed at $\pi/2$, the
rate for the diagonal and nondiagonal elements, $\sigma_{+-}/\sigma$, 
can achieve the smallest value -1.
\end{itemize}
We confirm that the above behaviors are reproduced in
the high $J/\psi$ momentum limit of the QCD $J/\psi\ccbar$
production cross section. Detailed studies of the inclusive
$J/\psi\ccbar$ cross section at the highest $J/\psi$ momentum
region may reveal the existence of an additional large
QCD enhancement factor in the exclusive limit of the process.
%
%
\section{Conclusions}
In this paper, we have calculated the inclusive $J/\psi$ productions in 
$e^+ e^-$ annihilation at the center-of-mass energy 10.58 GeV. Within the 
framework of NRQCD, we performed the leading-order calculations for the three 
production modes: the QCD and QED $J/\psi c{\bar c}$ processes and the 
$J/\psi gg$ process. The cross section ratio for the $J/\psi c {\bar c}$ and 
$J/\psi gg$ processes is found to be 1:1.5, larger than some previous 
estimates~\cite{cho,chao1,lee}. In order to explain the Belle experimental 
data~\cite{belle1,belle2,belle4}, we considered
the collinear Sudakov suppression for $J/\psi gg$ and a large renormalization
$K$ factor for the QCD $J/\psi c {\bar c}$ mode. We found that our predictions
with the $K$ factor of around 4 can reproduce both the inclusive $J/\psi$ 
momentum distribution and the large fraction of the $J/\psi\ccbar$ events
~\cite{belle1,belle2,belle4}. We have presented a complete density-matrix 
analysis and considered the diagonal elements as well as nondiagonal ones 
which give rise to polar and azimuthal angle correlations
in the $J/\psi$ leptonic decays. The various momentum and angular distributions
have been plotted, and their sensitivity to the magnitude of the $K$ factor 
has been studied.
Finally we briefly discussed the relation between the inclusive and exclusive 
processes.
\par
Throughout this work, we have not presented any discussion on the relativistic
corrections to the inclusive $J/\psi$ production. We know that the relativistic
corrections may play a very important role since the charm quark is not
so heavy and its velocity $v$ is not so small within the charmonium.
In fact, with such corrections, the cross section of the exclusive process
is doubled~\cite{braaten-lee}. In the case of
inclusive processes where only one charmonium is produced, the
relativistic corrections should not exceed that for the
exclusive one. Therefore if one knows how large the relativistic
corrections affect the total cross section, a significant fraction of the 
large $K$ factor may be explained.
%
%
\par
Before closing the discussion, we note here that we find no
evidence of the color-octet contribution to the prompt $J/\psi$
production process. If its contribution is significant,
the color-singlet contribution to the $J/\psi gg$ process
should be further suppressed in order to keep the large
fraction of the $J/\psi\ccbar$ process. There is a possibility
that the color-octet processes contribute mainly to the
$\psi'$ production and that its contribution to direct $J/\psi$
production is suppressed.  Detailed studies of prompt $J/\psi$
and $\psi'$ production data in e+e- B factories may reveal
the nature of charmonium production dynamics.
%
\vspace{.3cm}
\par
{\bf Acknowledgments}
\vspace{.2cm}

We would like to thank Bruce Yabsley for discussions about Belle data,
and Jun-ich Kanzaki for discussions about the BASES program.
ZHL and GHZ thank Ming-Xing Luo for numerous helpful discussions.

The works of KH, ZHL and GHZ are supported in part by Grant-in-Aid Scientific
Research from MEXT, Ministry of Education, Culture, Science and Technology of
Japan. The works of ZHL and GHZ are supported in part by the Japan Society for
the Promotion of Science (JSPS).
EK was supported by the Belgian
Federal Office for Scientific, Technical and Cultural Affairs through the
Interuniversity Attraction Pole P5/27.
CFQ is supported in part by the National Science Foundation of China.
\vspace{.3cm}
\par
\appendix{:~Helicity amplitude for NRQCD}
In this appendix, we present explicitly how we calculate the helicity 
amplitudes for NRQCD processes by using the HELAS subroutines~\cite{helas}.
\par
Consider the inclusive production
of a quarkonium $H$ with momentum $P$ and helicity $\lambda$ in colliders,
$a+b\to H_\lambda(P)+X$, where $a$ and $b$ stand for the initial particles
which can be $e^+ e^-$ or a parton pair and $X$ denotes additional final states.
If $H$ is a $S$-wave bound state, to the lowest order in the velocity $v$,
both quark and anti-quark within the quarkonium are on-shell and carry the 
momentum $P/2$. Hence the total amplitude can be separated into two 
independent parts as follows:
\bea\label{amp}
M(a+b\to H_\lambda(P)+X)&=&\sum_{\lambda_1,\lambda_2,i,j}
{\cal A}(a+b\to Q_{\lambda_1}^i(P/2)+{\bar Q}_{\lambda_2}^j(P/2) +X) \nonumber \\
& &\times {\cal F}(Q_{\lambda_1}^i(P/2)+{\bar Q}_{\lambda_2}^j(P/2)\to H_\lambda(P)),
\eea
where ${\cal A}$ is the helicity amplitude in the parton level and ${\cal F}$
is the amplitude for forming a bound state.
Here $Q$(${\bar Q}$) should be charm or bottom quark(anti-quark) with helicity
$\lambda_1$($\lambda_2$) and color $i(j)$.
The momenta and helicities for the other particles have
been suppressed in Eq. (\ref{amp}). More explicitly, ${\cal A}$ and ${\cal F}$
have forms
\bea\label{subamp}
{\cal A}&=&{\bar u}_{\lambda_1}^i(P/2) \Gamma v_{\lambda_2}^j(P/2), \nonumber\\
{\cal F}&=&{\bar v}_{\lambda_2}^j(P/2) \frac{N}{\sqrt{3}}
(\gamma_5 \alpha+{\slash \epsilon}_\lambda^*(P)\beta) u_{\lambda_1}^i(P/2).
\eea
Here $\Gamma$ depends on various production processes and
has a complicated structure which consists of the wave functions of initial
particles $a$ and $b$ as well as the final particles in $X$.
On the other hand ${\cal F}$ has a simple structure where $\alpha=1$,$\beta=0$
for pseudo-scalars and $\alpha=0$,$\beta=1$ for vectors.
$N$ is a normalization constant related to the meson wave function at the origin
$\Psi(0)=R(0)/\sqrt{4\pi}$.
The traditional definition of $HQ{\bar Q}$ vertex is given by\cite{keung-kuhn}
\bea
\frac{\Psi(0)}{2\sqrt{2m}}\frac{\delta_{ij}}{\sqrt{3}}
{\slash \epsilon}^*({\slash P}+2m).
\eea
where $m$ is the heavy quark mass.
The above vertex can be obtained identically by contracting
${\bar u}_{\lambda_1}(P/2)$ and $v_{\lambda_2}(P/2)$ in ${\cal A}$
with ${\cal F}$ and summing over the helicities $\lambda_1$ and $\lambda_2$.
After simplification of the Dirac matrices, one can find a simple
relation between the normalization constant and the wave function
\bea{\label{N}}
N=\frac{\Psi(0)}{(2m)^{3/2}}.
\eea
\par
Both of the amplitudes ${\cal A}$ and ${\cal F}$ in Eq. (\ref{subamp})
can be expressed in a transparent manner by using HELAS subroutines.
In particular, the program MadGraph can automatically generate
a Fortran code for ${\cal A}$ such that we only need to write a code
for ${\cal F}$ and combine ${\cal A}$ and ${\cal F}$ correctly.
After summing over the helicities for the intermediate heavy quark
and anti-quark as depicted in Eq. (\ref{amp}), we obtain
the full amplitude $M_\lambda$ for different polarized quarkoniums.
This $M_\lambda$ is what we need for calculating the spin density
matrix in Sec. 4.
\par
The technique can be utilized also for the quarkonium decay, for instance 
$J/\psi \to e^+ e^-$. The amplitude has the form
\bea\label{decayamp}
M(J/\psi_\lambda(P)\to e^+ e^-)&=&\sum_{\lambda_1,\lambda_2,i,j}
{\cal F}(J/\psi_\lambda(P)\to
c_{\lambda_1}^i(P/2)~{\bar c}_{\lambda_2}^j(P/2)), \nonumber \\
& & \times
{\cal A}(c_{\lambda_1}^i(P/2)~{\bar c}_{\lambda_2}^j(P/2) \to e^+ e^-)
\eea
with
\bea\label{decaysubamp}
{\cal F}&=&{\bar u}_{\lambda_1}^i(P/2) \frac{N}{\sqrt{3}}
{\slash \epsilon}_\lambda(P) v_{\lambda_2}^j(P/2),\nonumber\\
{\cal A}&=&{\bar v}_{\lambda_2}^j(P/2) \Gamma' u_{\lambda_i}^j(P/2).
\eea
With this amplitude, one can easily obtain the leptonic decay width.
By comparing with the analytic formula for the leptonic decay,
we verify that our normalization factor $N$ in Eq. (\ref{N}) is correct.
\par
We can also apply the method to calculate the exclusive processes
where the additional heavy quark pair should form another quarkonium.
The amplitude is expressed as
\bea\label{examp}
& &M(a+b\to H_1^\lambda(P_1)+H_2^{\lambda'}(P_2)) \nonumber \\
&=&\sum_{\lambda_1,\lambda_2,i,j}
\sum_{\lambda_3,\lambda_4,k,l}
{\cal A}(a+b\to Q_{\lambda_1}^i(P_1/2)+{\bar Q}_{\lambda_2}^j(P_1/2)
+Q_{\lambda_3}^k(P_2/2)+{\bar Q}_{\lambda_4}^l(P_2/2)) \nonumber \\
& &\times {\cal F}_1(Q_{\lambda_1}^i(P_1/2)+{\bar Q}_{\lambda_2}^j(P_1/2)
\to H_1^\lambda(P_1)) \nonumber \\
& &\times {\cal F}_2(Q_{\lambda_3}^k(P_2/2)+{\bar Q}_{\lambda_4}^l(P_2/2)
\to H_2^{\lambda'}(P_2)),
\eea
where the production amplitude is now expressed as
\bea\label{exsubamp}
{\cal A}&=&{\bar u}_{\lambda_1}^i(P_1/2) \Gamma_1 v_{\lambda_2}^j(P_1/2)
{\bar u}_{\lambda_3}^k(P_2/2) \Gamma_2 v_{\lambda_4}^l(P_2/2)
\eea
and ${\cal F}_1$ and ${\cal F}_2$ have similar structures as those
in Eq. (\ref{subamp}) except for different momentum, helicity and color
indices. Summation over the helicities and color
indices of two pairs of $c$ and ${\bar c}$ wave functions gives
the helicity amplitudes for the exclusive process $e^+e^-\to
J/\psi \eta_c$.

\end{document}